\documentclass[a4paper]{panl}
\usepackage{cite}
\usepackage{wrapfig}
\usepackage{graphicx}
\usepackage{amssymb}
\usepackage{amsfonts}
\usepackage{amsmath}
\usepackage{longtable}
\usepackage{rotating}
\usepackage{lscape}
\usepackage{epsfig}
\usepackage{multirow}
\usepackage{orcidlink}

\usepackage{xcolor}

\originalTeX
\begin{document}

\title{What do we know
about the confinement mechanism?
}
\maketitle
\authors{
Zeinab Dehghan \orcidlink{0000-0002-5976-9231}, $^{a,}$\footnote{E-mail: zeinab.dehghan@ut.ac.ir},
Manfried Faber
\orcidlink{0000-0002-3572-5429},
$^{b,}$\footnote{E-mail: faber@kph.tuwien.ac.at}	
}
\setcounter{footnote}{0}
\from{$^{a}$\,Department of Physics, University of Tehran, Tehran 14395-547, Iran}
\from{$^{b}$\,Atominstitut, Technische Universität Wien}

\begin{abstract}

\
\vspace{0.2cm}


Color confinement is a fundamental phenomenon in quantum chromodynamics. In this work, the mechanisms underlying color confinement are investigated in detail, with a particular focus on the role of non-perturbative phenomena such as center vortices and monopoles in the QCD vacuum. By exploring lattice QCD approaches, including the Maximal Center Gauge and center projection methods, we examine how these topological structures contribute to the confining force between color charges. We also address the limitations of conventional methods and suggest improvements to the gauge fixing prescription to enhance the accuracy of string tension predictions. Our findings support the validity of the center vortex model as a key candidate for understanding the dynamics of the confining QCD vacuum.

\end{abstract}
\vspace*{6pt}


\label{sec:intro}
\section{Introduction}

The phenomenon of color confinement, a cornerstone of quantum chromodynamics (QCD), ensures that quarks and gluons are perpetually bound within hadrons. This paper provides a review of the current understanding of the confinement mechanism. Lattice QCD simulations provide strong numerical evidence supporting these models. Experimental observations from the GlueX experiment further substantiate the confinement hypothesis by providing detailed insights into the spectrum of light mesons and the search for hybrid mesons~\cite{Somov:2015qht, Dobbs:2019sgr}.

Various theoretical models have been proposed to explain the confinement mechanism. One of the most prominent is the dual superconductor model, introduced independently by ’t Hooft and Mandelstam, which suggests that the QCD vacuum behaves like a dual type-II superconductor, with magnetic monopole condensation leading to confinement through the formation of color-electric flux tubes~\cite{Mandelstam:1974pi, tHooft:1977nqb} between electric color charges. Kronfeld, Schierholz, and Wiese demonstrated the importance of abelian monopoles in the confinement process through the Maximal Abelian Gauge (MAG) ~\cite{Kronfeld:1987ri}.  Another important approach is the center vortex model, which attributes confinement to the formation of closed magnetic flux tubes—vortices—that are quantized to center elements of the gauge group~\cite{DelDebbio:1996lih, Faber:1997rp}. These vortices are believed to form a complex network that permeates the QCD vacuum, contributing to the area-law behavior of Wilson loops, a signature of confinement. Abelian monopoles and center vortices have been identified as key players in the dynamics of confinement. Lattice simulations have revealed the presence of these topological defects in the QCD vacuum, with numerical evidence supporting their role in generating the string tension observed in the potential between quarks~\cite{Bali:1994de, Greensite:2003bk}.

One of the primary tools for investigating and detecting center vortices in lattice QCD is the Maximal Center Gauge (MCG), a gauge fixing procedure that projects the gauge field onto the center elements of the SU(N) group~\cite{DelDebbio:1998luz}. Despite its success in identifying vortices, the MCG method faces challenges, particularly due to the Gribov ambiguity, which arises from multiple local maxima in the gauge fixing process, complicating the detection of physically meaningful vortices.  In this paper, we investigate the various models and mechanisms that contribute to our understanding of color confinement, with a particular focus on the role of center vortices. Additionally, we propose refinements to existing methods to address current limitations in vortex detection and string tension calculations.

\label{sec:topological defects}
\section{Monopoles, vortices and confinement}

Magnetic monopoles provide a promising model for confinement through the dual superconductor analogy. In the dual superconductor model, magnetic monopole condensation creates a phase where the color-electric field between quarks is squeezed into narrow tubes, similar to the Meissner effect in superconductivity, where magnetic fields are expelled. The field lines between a quark and an antiquark form a flux tube, whose tension is independent of the distance, leading to confinement. This constant string tension results in a linear potential between quarks and antiquarks. By using lattice simulations in the maximally Abelian gauge, it has been shown that monopole density correlates strongly with confinement regions, suggesting a critical role for monopoles in creating the necessary flux tubes. Additionally, removing monopole contributions from simulations disrupts confinement, indicating that monopoles are essential for maintaining the confining properties of the QCD vacuum.

Center vortices are topological field configurations, quantized magnetic flux tubes, which form closed two-dimensional surfaces in dual space-time. If they penetrate surfaces limited by Wilson loops, they contribute to the Wilson loop with a center element of the respective gauge group, i.e. with an Abelian factor. The area law of Wilson loops and thus confinement results from the quantum fluctuations in the number of center vertices linking the loop. Vortices are thick flux tubes, of unlimited size, which fill the vacuum in the confinement phase.  In lattice calculations, the thick vortices are usually identified by transforming the link variables as close as possible to center elements and then projecting them onto center elements. This creates infinitely thin surfaces from projected vortices (P-vortices). Notably, when center vortices are removed from the lattice, confinement effects disappear, indicating that these structures are essential for the confinement mechanism.

\begin{figure}[h!]
\centering
(a)\includegraphics[scale=0.25]{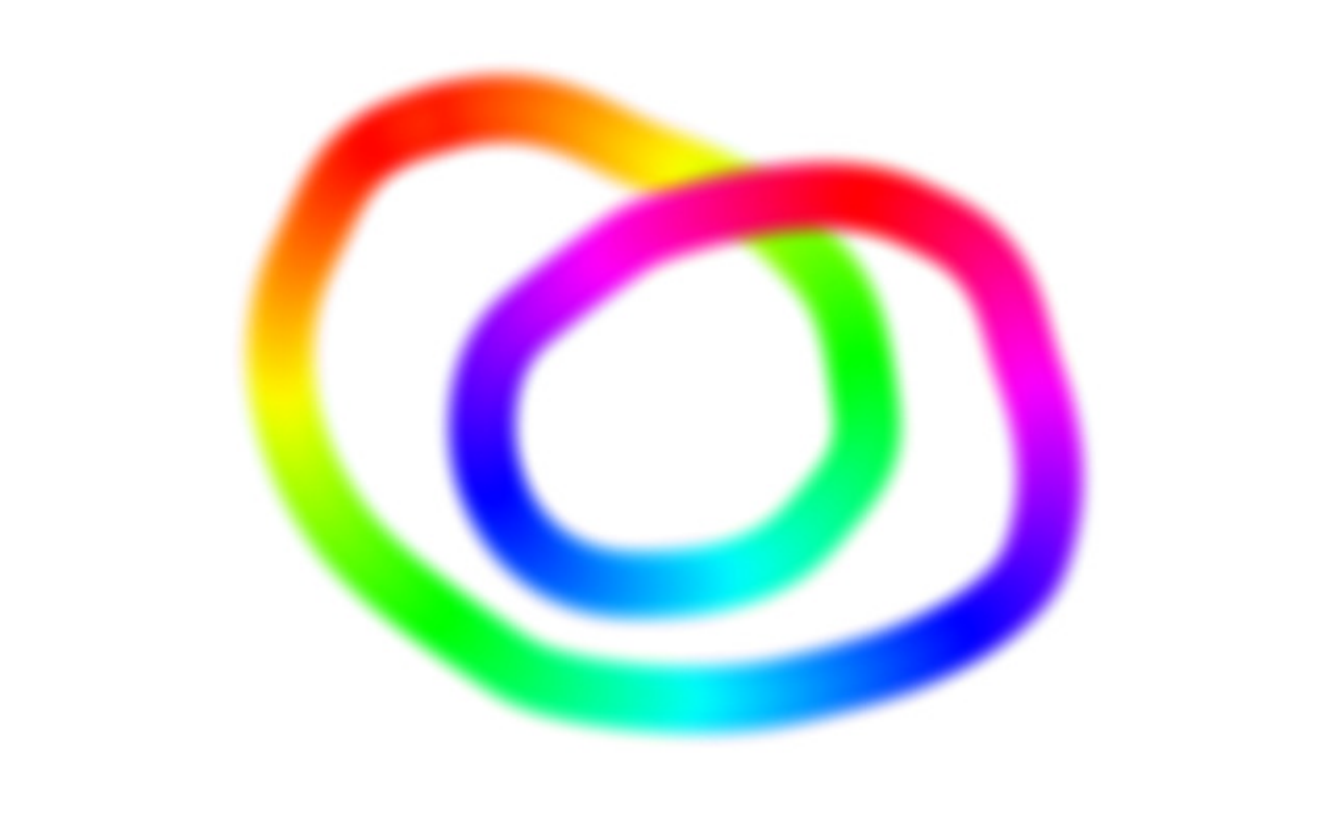}
\hspace{50mm}
(b)\hspace{-40mm}\includegraphics[scale=0.7]{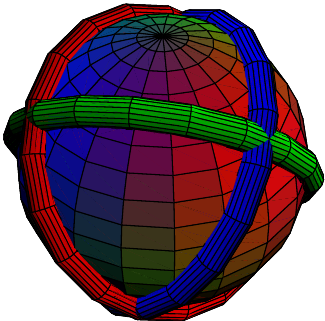}
\caption{\small a) A schematic diagram of a colorful vortex. Observing such a vortex with a red color filter one finds sources of red color. b) Abelian projection of the spherical vortex yields monopole lines (circles) depending on the U(1) subgroup~\cite{Faber:2007cov}.}
\label{fig:chain}
\end{figure}
Vortices are colorful, in 3 dimension closed colorfull lines. In the adjoint representation of SU(2), they have three colors, see Fig.~\ref{fig:chain} (a). Abelian projection observes the color flux with the color filter of the respective U(1) subgroup. With a red filter the vortex contains sources and sinks of red color flux, i.e. red monopoles. In 4 dimensions a thick colorful symmetric spherical vortex, which is a non-Abelian vortex, defines a surface, see Fig.~\ref{fig:chain} (b). After Abelian projection it shows different positions of the monopole world lines, depending on the projection to the $\sigma_1, \sigma_2$ and $\sigma_3$ directions.

\label{sec:mcg modification}
\section{A new perspective on the MCG definition}

Assuming center excitations are essential for confinement, we identify center vortices in the lattice by employing the Wilson action for SU(2) gauge theory. Center vortices are usually detected in direct maximal center gauge (MCG) where a gauge transformation $g(x) \in $ SU(2) is applied to shift the link variables $U_\mu(x)$ as close as possible to center elements by maximizing the gauge functional
\begin{equation}\label{eq:GaugeFunctR}
R_\text{MCG}=\sum_x\sum_\mu\mid\text{Tr}[{}^gU_\mu(x)]\mid^2,
\end{equation}
with ${}^gU_\mu(x)=g(x+\hat\mu)\,U_\mu(x)\,g^\dagger(x)$. Then the transformed link variables ${}^gU_\mu(x)$ are projected onto the nearest center element, which is $\pm 1$ for SU(2),
\begin{equation}\label{eq:CentrProj}
{}^gU_\mu(x)\rightarrow Z_\mu(x)\equiv \mathrm{sign Tr}[{}^gU_\mu(x)].
\end{equation}
Non-trivial center projected plaquettes $Z_{\mu\nu}(x)=-1$, known as P-plaquettes, contain one or three non-trivial links. On a four-dimensional lattice, each link belongs to six plaquettes. On the dual lattice, the six P-plaquettes form the surface of a cube. It turns out that in the confinement phase, a large number of negative links are arranged in such a way that the dual P-plaquettes form extended closed surfaces: vortices percolating through the whole space-time lattice.

Finding the global maximum of the gauge functional~(\ref{eq:GaugeFunctR}) is a non-polyno\-mial (NP) problem. This gauge functional includes numerous local maxima that can be approached using the conjugate gradient method, with the global maximum being approximated incrementally. Engelhardt and Reinhardt~\cite{Engelhardt:1999xw} have analytically shown that a trivial maximization cannot accurately yield the correct physics. Refs.~\cite{Kovacs:1998xm, Bornyakov:2000ig} have demonstrated numerically that increasing the gauge functional values reduces the string tension. Despite this property, the ensemble of local maxima of the gauge functional contains the information about the string tension. To show this, we investigate the distribution of these local maxima, extract the string tensions from the configurations as a function of the maxima of the gauge functional and modify the prescription for the maximization of the functional.

\begin{figure}[t]
\includegraphics[scale=0.75]{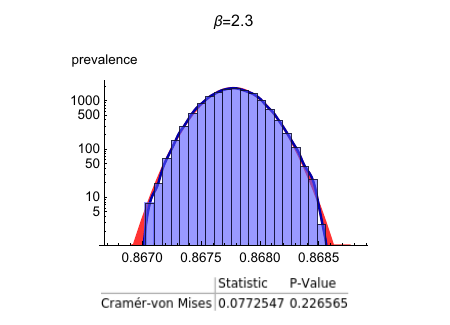}
\hspace{-20mm}
\includegraphics[scale=0.75]{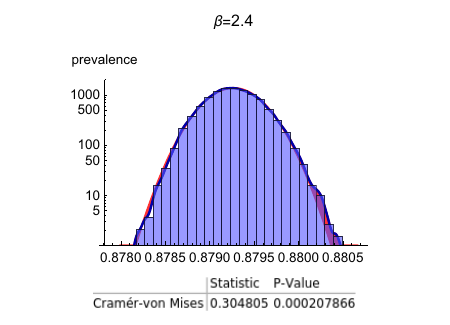}
\hspace{-20mm}
\includegraphics[scale=0.75]{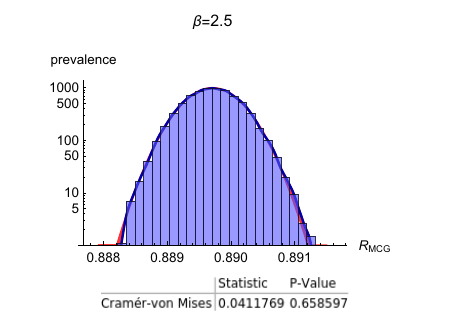}\\
\includegraphics[scale=0.75]{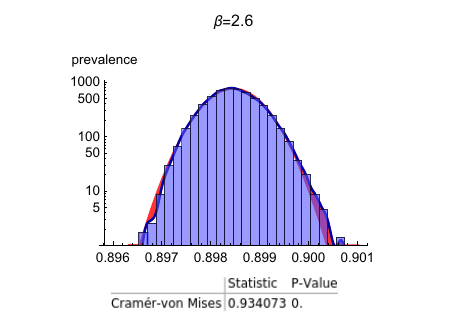}
\hspace{-20mm}
\includegraphics[scale=0.75]{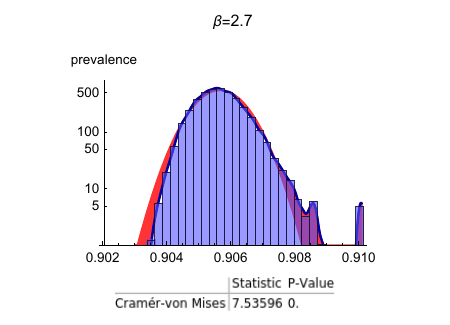}
\hspace{-10mm}
\raisebox{7mm}{\includegraphics[scale=0.55]{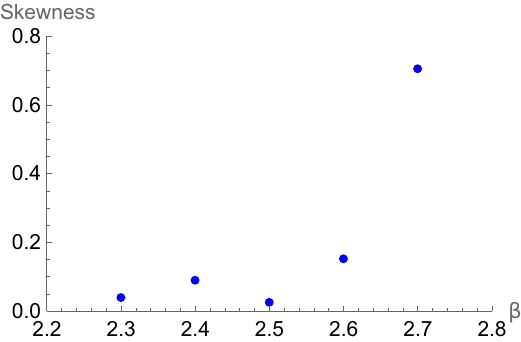}}
\caption{\small For the lattices with $\beta=2.3$ to $2.7$, the distributions of 20\,000 local gauge maxima of $R_\text{MCG}$ values are shown on a logarithmic scale. The blue solid lines represent smooth kernels, with the histogram of the data in the background. Red highlights indicate deviations from the normal distribution. The results of distribution tests are provided below the distribution. The last plot illustrates the variation in skewness of local gauge maxima distributions for various $\beta$ values.}
\label{fig:DistriDistortion}
\end{figure}

On a lattice of size $24^4$ for $\beta$ values $2.3, 2.4, 2.5, 2.6$ and $2.7$, we employed Monte-Carlo simulation with 20 random starts, 3000 initial sweeps and generated 10 configurations separated by 1000 sweeps, resulting in 200 configurations. For each configuration, 100 unbiased random gauge copies were performed, yielding 20\,000 gauge configurations for each $\beta$. In Fig.~\ref{fig:DistriDistortion}, for $\beta=2.3$ to $2.7$ we show the logarithmic plots of the distributions obtained from these 20\,000 gauge fields at the local maxima of $R_\text{MCG}$, with bin widths set to one-third of the standard deviation. The full blue lines represent a smooth kernel distribution, which provides a continuous substitute for histograms. We find that these distributions are very close to symmetric Gaussian distributions. The red areas indicate the deviations from the Gaussian fit. Below the distribution, we show the Cram$\acute e$r–von Mises criterion values, which are used to evaluate the goodness of fit to the normal Gaussian distribution. The p-values under $0.05$ denote statistically significant deviations from normality. The skewness, defined as the ratio of the third central moment to the $3/2$ power of the second central moment, quantifies deviations from symmetry in a distribution and is shown in the final plot in Fig.~\ref{fig:DistriDistortion}. For $\beta=2.3, 2.4$ and $2.5$ the distributions of the local gauge maxima are nearly symmetric and Gaussian. For $\beta = 2.6$ we observe the first systematic deviations from the normal distribution and for $\beta=2.7$ the deviation becomes very noticeable.

\begin{figure}[t]
\includegraphics[scale=0.4]{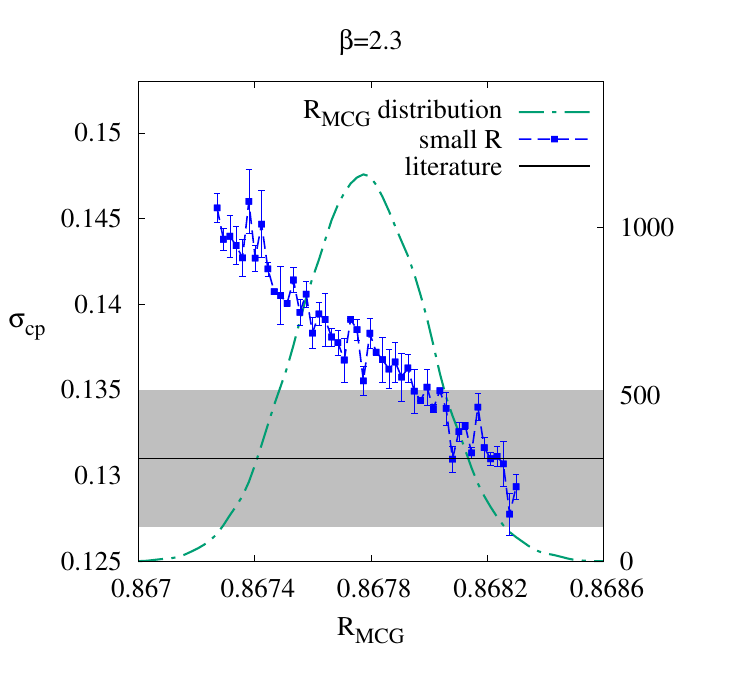}
\hspace{-10mm}
\includegraphics[scale=0.4]{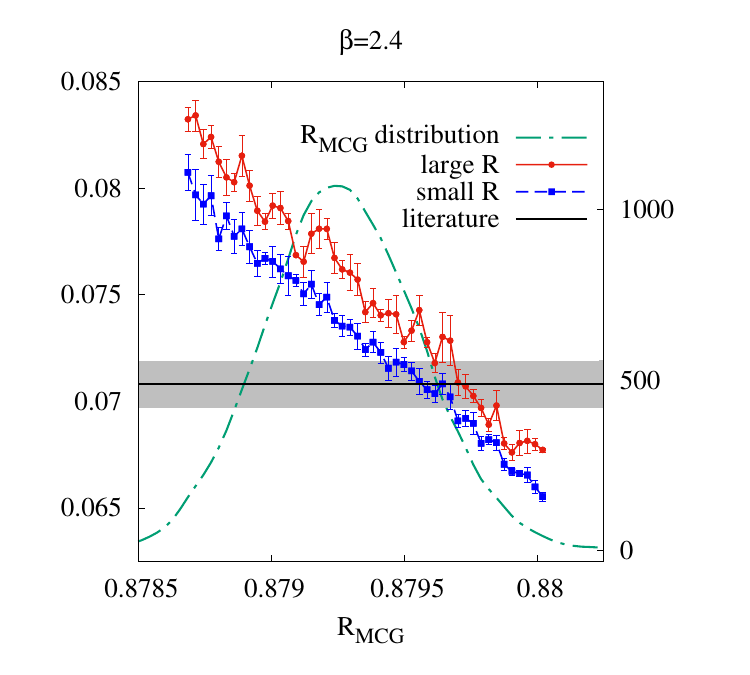}
\hspace{-10mm}
\includegraphics[scale=0.4]{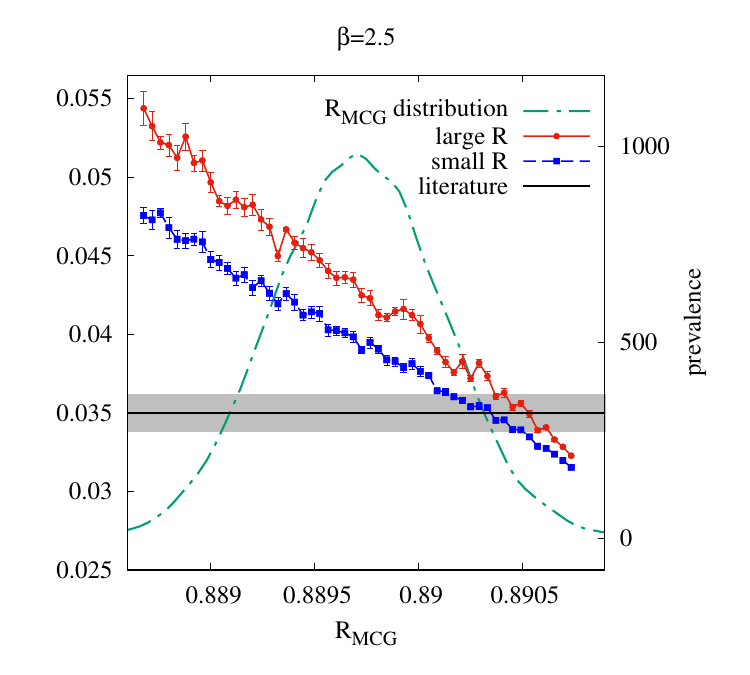}
\caption{\small
For lattices with $\beta=\in \{2.3, 2.4, 2.5\}$, the center-projected string tensions, derived from 200 configurations around specified $R_\mathrm{MCG}$  values, are compared to the unprojected string tensions reported in the literature, represented by horizontal black lines~\cite{Bali:1994de, Caselle:2015tza, Fingberg:1992ju, Michael:1990fh}. The Gaussian distributions of the corresponding local gauge maxima are indicated by dash-dotted lines. The intersection of the red data with the literature values, located at the right tip of the distributions, reveals the best agreement of $\sigma_\mathrm{cp}$ with the asymptotic string tensions.}
\label{fig:bothsigmas}
\end{figure}

\begin{figure}[ht]
\raisebox{1.4mm}{
\includegraphics[scale=0.34]{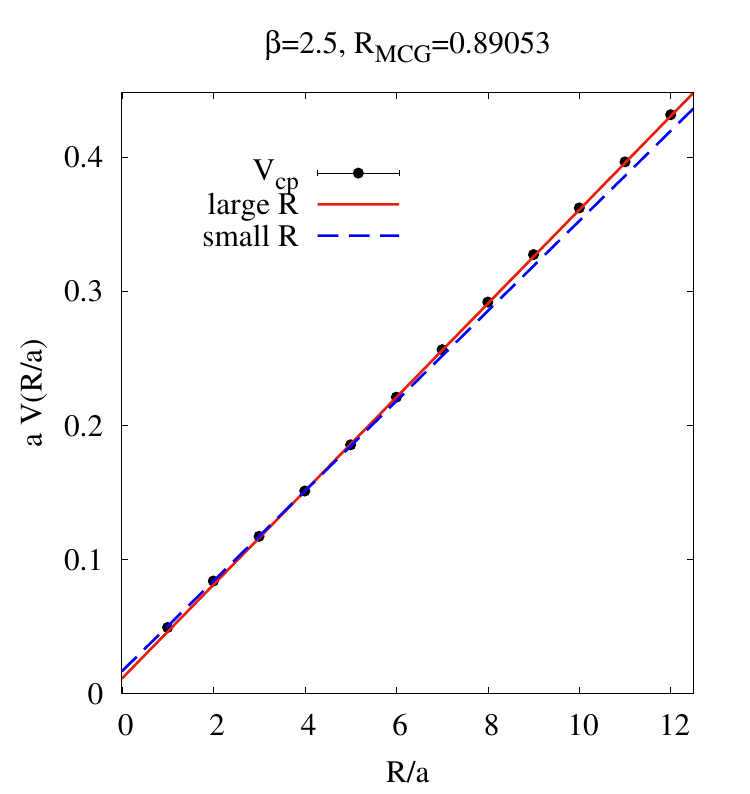}}
\hspace{-5mm}
\includegraphics[scale=0.41]{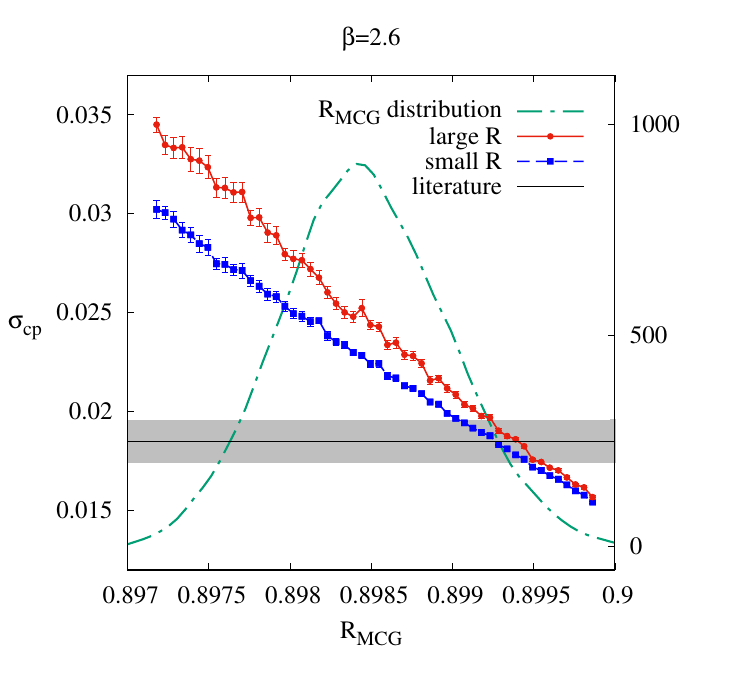}
\hspace{-12mm}
\includegraphics[scale=0.41]{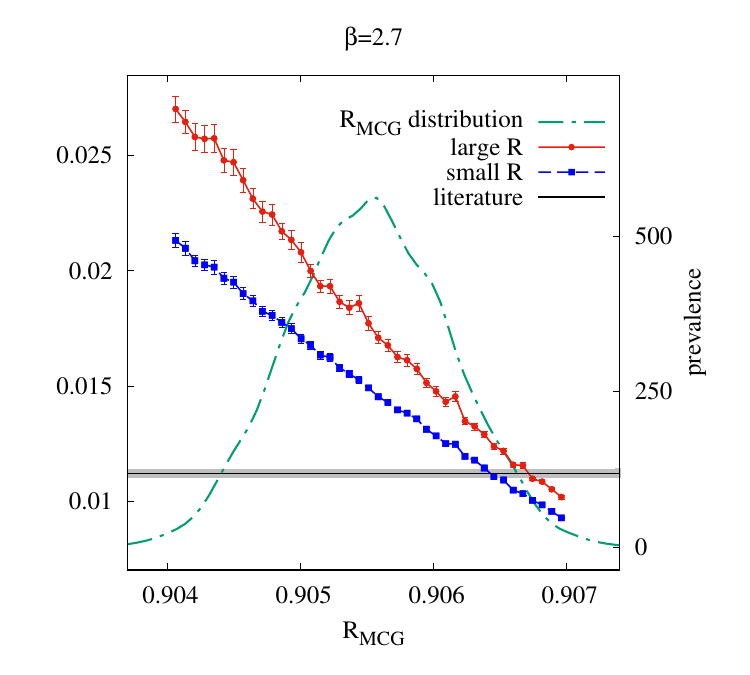}
\caption{\small
As an example, for the lattice with $\beta=2.5$, the first plot shows the potential $V_\mathrm{cp}(R)$ extracted from 200 configurations centered around $R_\text{MCG}=0.89053$. The two other diagrams are analogous to the diagrams in Fig.~\ref{fig:bothsigmas}, but for higher $\beta$ values, where the ensembles of local maxima were symmetrized for $\beta = 2.6$ and corrected as well as symmetrized for $\beta = 2.7$.}
\label{fig:bothsigmas2}
\end{figure}

We define the static potential and, consequently, the asymptotic slope of the potential, $\sigma_\mathrm{cp}$, from the expectation values of the center projected Wilson loops $W_\mathrm{cp}(R,T)$. Fig.~\ref{fig:bothsigmas} shows $\sigma_\mathrm{cp}(R_\mathrm{MCG})$ for lattices with symmetric Gaussian distributions, i.e., for $\beta \in \{2.3, 2.4, 2.5\}$, where the string tensions are extracted from 200 configurations around the specified values of $R_\mathrm{MCG}$. The slopes of the potential at small and large distances are represented by blue and red colors, respectively. This figure exhibits an almost linear relationship between the string tension and the gauge functional. We also found the same nearly linear behavior for Creutz ratios, see Ref.~\cite{Dehghan:2023gep}. The horizontal black lines in Fig.~\ref{fig:bothsigmas} represent the string tensions extracted from unprojected configurations~\cite{Bali:1994de, Caselle:2015tza, Fingberg:1992ju, Michael:1990fh}. The normalized distributions of the $R_\mathrm{MCG}$ values are shown by dashed-dotted lines. It is observed that an extreme value of the local maxima of the gauge functional results in an underestimate of the string tension. The string tension, the slope of potential at large distances, extracted from center projected configurations approaches the string tension of unprojected configurations at the right tip of the distribution, for $\beta=2.5$ around $R_\mathrm{MCG}=0.8905$. For the evaluation of the string tensions in Fig.~\ref{fig:bothsigmas}, it is noteworthy that the potentials $V(R)$ show a precautious linearity, of which the left diagram in Fig.~\ref{fig:bothsigmas2} gives an example. Nevertheless, there is generally a small kink in the linearity of the potential, so that for $\beta\ge2.4$ it makes sense to specify two string tension for smaller and larger $R$, see Fig.~\ref{fig:bothsigmas}. For $\beta=2.3$, however, the quark-antiquark potential $aV(R/a)$ increases rapidly due to the large lattice constant $a$ and the signal disappears to early in the noise and only one value of the string tension can be determined. The precautious linearity of $V(R)$, as shown in the left diagram of Fig.~\ref{fig:bothsigmas2}, can be explained by closed vortices that percolate the lattice and randomly pierce the minimal area of the Wilson loops. Short-range fluctuations of the vortex surfaces increase the average number of piercings for small loops. If the size of the Wilson loops exceeds the extent of these fluctuations, the piercings become correlated and no longer contribute to the string tension. The slight kink in the potentials therefore provides information about the maximal size of the short-range fluctuations.

For $\beta = 2.6$, where we observed in Fig.~\ref{fig:DistriDistortion} first deviations from the normal Gaussian distribution, it is sufficient to symmetrize the ensembles of local maxima as described in Ref.~\cite{Dehghan:2024rly}. The second plot in Fig.~\ref{fig:bothsigmas2} shows the resulting $\sigma_\mathrm{cp}(R_\mathrm{MCG})$ for the remaining configurations after symmetrization. For $\beta = 2.7$, for which we have already found strong deviations from the Gaussian distribution in Fig.~\ref{fig:DistriDistortion}, we can either accept that this $\beta$ lies outside the window in which MCG works, or better try to determine the cause of the deviations. A detailed examination of the $V(R)$ curves shows that for high values of $R_\mathrm{MCG}$ a deviation from the previously established $R$-dependence of the potentials occurs with increasing frequency; the string tension for high $R$ falls below the string tension for low $R$. Its reason lies in the incorrect detection of vortices, which leads to P-vortices with finite extension. This reversal of the behavior of the projected string tension is now used as a criterion to exclude gauge copies from the averaging. The remaining copies, which are corrected and symmetrized, result in the last diagram in Fig.~\ref{fig:bothsigmas2} and show a behavior that is consistent with the other $\beta$ values. The gauge copies in the right-hand tips of the Gaussian distributions shown in Fig.~\ref{fig:bothsigmas2} reproduce the literature values of the string tensions just as well as those in Fig.~\ref{fig:bothsigmas}.

\begin{figure}[ht]
\centering
\includegraphics[scale=0.37]{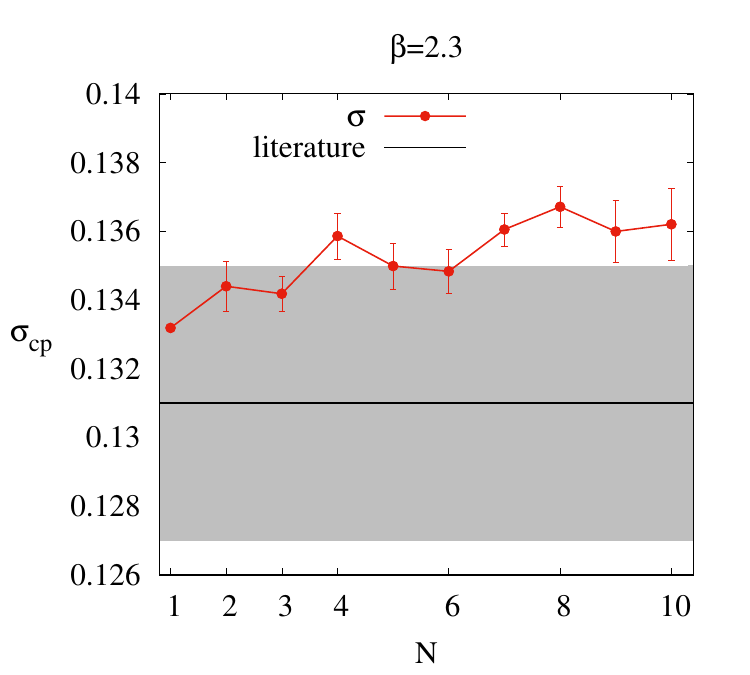}
\hspace{-5mm}
\includegraphics[scale=0.37]{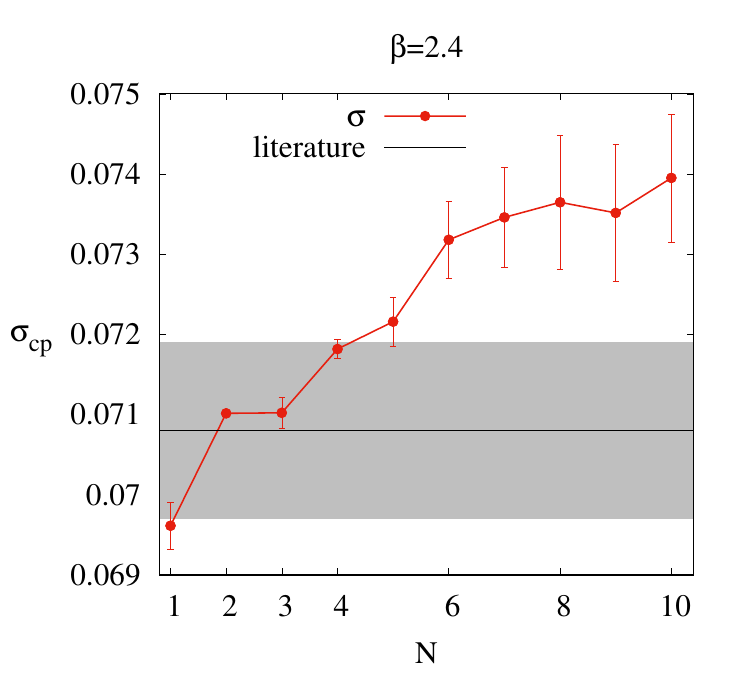}
\hspace{-5mm}
\includegraphics[scale=0.37]{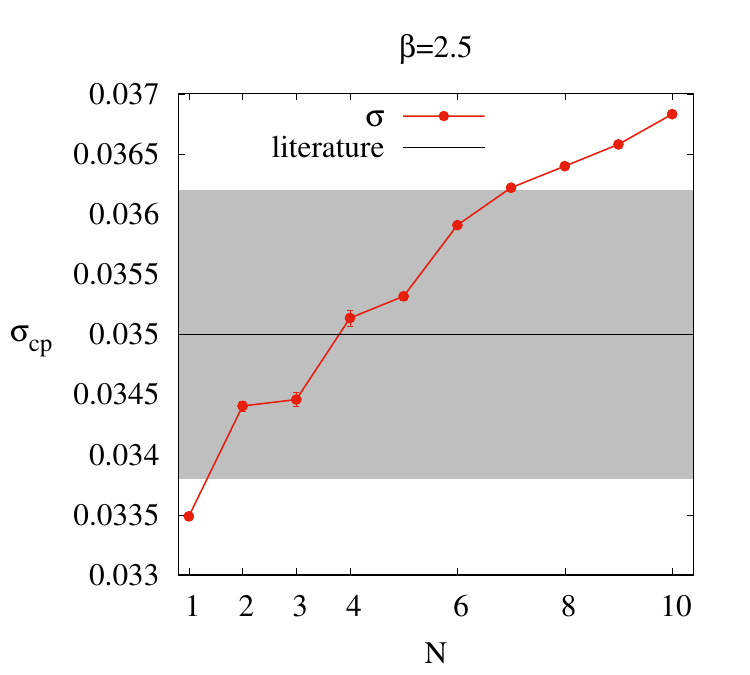}
\\
\includegraphics[scale=0.37]{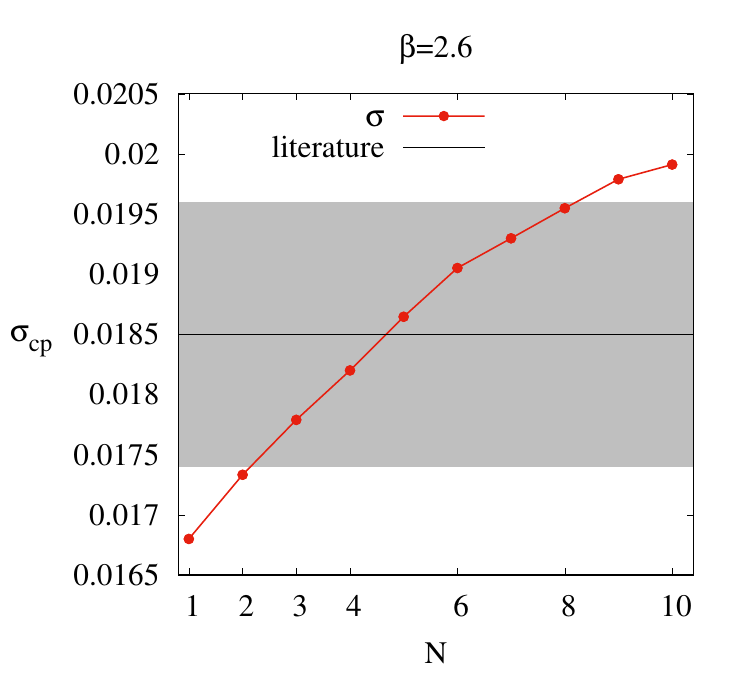}
\hspace{-5mm}
\includegraphics[scale=0.37]{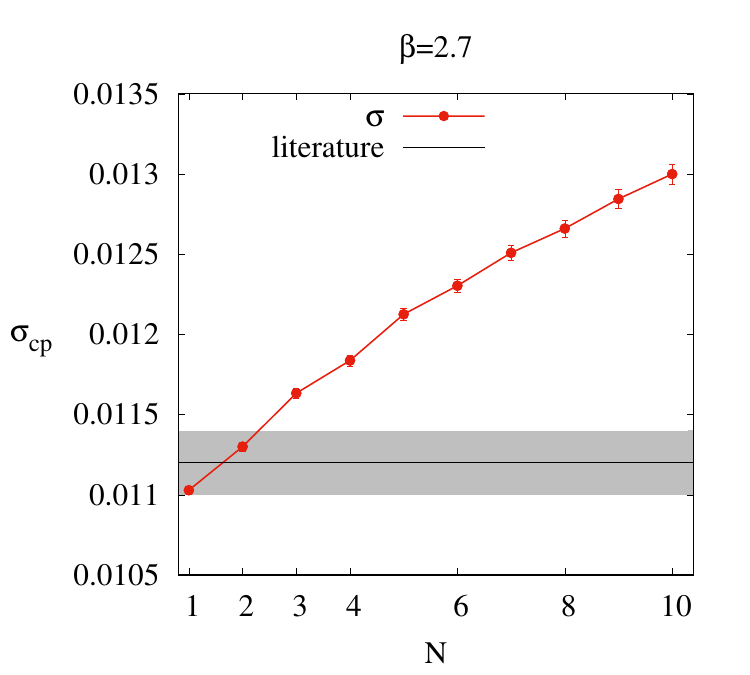}
\caption{\small For lattices with $\beta=2.3$ to $2.7$, center projected string tensions are extracted from 200 physical fields from the $N$ gauge copies with the highest values of $R_\mathrm{MCG}$ within the Gaussian distributed subset of the ensemble of random gauge copies. . The horizontal black lines represent string tensions obtained from unprojected configurations in Refs.~\cite{Bali:1994de, Caselle:2015tza, Fingberg:1992ju, Michael:1990fh}.}
\label{fig:SN}
\end{figure}

Encouraged by these results, in Fig.~\ref{fig:SN} we systematically investigate the accuracy with which the gauge configurations with the highest $R_{MCG}$ values within these Gaussian distributions correctly predicts the string tensions. For this purpose, we evaluate the $N$ gauge copies with the highest $R_{MCG}$ values within the Gaussian distributions for each physical field configuration. For $N=1$, the highest $R_\mathrm{MCG}$ values underestimate the string tension. Since $R_\mathrm{MCG}$ and $\sigma_\mathrm{cp}$ are approximately linearly related, an increase in $N$ leads to an increase in string tension. In general, we observe good agreement between center projected and unprojected string tensions with two or three gauge copies. For $N > 7$, the string tensions are overestimated. Consequently, unrestricted maximization, which tends to underestimate the string tension, can be effectively replaced by a maximization restricted to the Gaussian distributed subset of the ensemble of random gauge copies.

\label{sec:conclusion}
\section{Conclusion}

The confinement mechanism is one of the most challenging and fascinating aspects of QCD and particle physics. Despite notable progress in theoretical models and computational simulations, confinement is still an active field of research and debate. Confinement is based on the non-triviality of the QCD vacuum, which is filled with quantum fluctuations. The dual superconductivity model and the vortex model answer the question of which of the possible quantum fluctuations are decisive for confinement. Both methods are based on gauges that emphasize the relevant degrees of freedom of the gauge field by maximizing a gauge functional. The maximal Abelian gauge functional~\cite{Kronfeld:1987ri} determines the world lines of magnetic monopoles in an Abelian subgroup of gauge degrees of freedom, while the maximal center gauge (MCG) functional~(\ref{eq:GaugeFunctR}) detects quantized magnetic flux tubes and approximates them by world surfaces of thin magnetic fluxes. The two mechanisms fit together well, since the world lines of the magnetic monopoles lie on the world surfaces of the vortices and reflect their color structure.

A discussion of the maximization condition in the MCG turned out to be necessary. The previous MCG procedure~\cite{DelDebbio:1996lih, Faber:1997rp} had predicted good values of the string tension. In this method, a few (e.g. 6) random gauge copies were selected and the best of these gauge configurations was chosen to determine the string tension. The investigations in~\cite{Kovacs:1998xm, Bornyakov:2000ig} made it clear that a targeted intensive search for the global maximum of the gauge functional is not suitable for determining the position of the vortices in a field configuration. As Engelhardt and Reinhardt~\cite{Engelhardt:1999xw} had already stated, the physically thick vortices cannot be approximated well by the thin, singular projected vortices. However, the successes of MCG  indicated that random gauge copies contain the full information about the physical string tension. We have confirmed this assumption in this work by showing that in a wide range of coupling strengths $2.3\le\beta\le2.6$ the $R_\mathrm{MCG}$ values of random copies are Gaussian distributed in good approximation and the maximization must be restricted to these Gaussian distributions. The maximization of the MCG gauge functional~(\ref{eq:GaugeFunctR}) within these random Gaussian distributions provide very good predictions of the string tensions. We have also shown that in the range around $\beta=2.7$ it is easy to see from the values of the projected Wilson loops which maximizations of the gauge functional fail. By excluding these gauge copies from the maximization, the string tension could also be reproduced well in this $\beta$ region.

Finally, we would like to mention that an effective model of color confinement would be desirable as further confirmation of the vortex model. Such considerations were initiated in~\cite{Oxman:2018dzp, Junior:2024urr}





\end{document}